\documentstyle[12pt]{article}

\begin{document}

\begin{titlepage}

\title{Bell-Kochen-Specker theorem:\\
A proof with 18 vectors\footnote{Phys. Lett. A 
{\bf 212}, 183 (1996).}}

\author{Ad\'{a}n Cabello\thanks{Electronic address: fite1z1@sis.ucm.es}\\
Jos\'{e} M. Estebaranz\\
Guillermo Garc\'{\i}a-Alcaine\\
{\em Departamento de F\'{\i}sica Te\'{o}rica,}\\
{\em Universidad Complutense, 28040 Madrid, Spain.}}

\date{January 29, 1996}

\maketitle

\begin{abstract}
We present a ``state-independent'' proof of the Bell-Kochen-Specker theorem
using only 18 four-dimensional vectors, which is a record for this kind of
proof. This set of vectors contains subsets which allow us to develop a
``state-specific'' proof with 10 vectors (also a record) and a
``probabilistic'' proof with 7 vectors which reflects the algebraic
structure of Hardy's nonlocality theorem.

\end{abstract}

\end{titlepage}

The Bell-Kochen-Specker (BKS) theorem \cite{1,2} asserts that there is no
consistent way of ascribing non-contextual definite answers to some sets of
yes-no questions regarding an individual physical system. There are
different versions of the theorem: ``state-independent'',
``state-specific'', and ``probabilistic'' \cite{3,4}.

Since the original state-independent proof by Kochen and Specker involving
projectors over 117 three-dimensional real vector was formulated, successive
demonstrations have reduced the size of the set to only 20 four-dimensional
vectors \cite{5}; see for instance the references in \cite{3,4}.

In this paper we present a state-independent proof with only 18 real vectors
in dimension 4. We also find subsets with 10 and 7 vectors making possible
state-specific and probabilistic proofs, respectively. Finally, we show the
relation between our probabilistic proof and a theorem by Hardy \cite{6} on
the incompatibility between quantum mechanics and local realistic theories.

Given an individual physical system, let $v({\bf u})$ denote the answer ($1=%
{\rm yes}$, $0={\rm no}$) in said system to the proposition $P_{{\bf u}}$
(mathematically represented by the projector $\left| {\bf u}\right\rangle
\left\langle {\bf u}\right| $) in a non-contextual hidden-variables (NCHV)
theory. In order to simplify the notation we will write ${\bf u}$ as a row
vector, omit its normalization constant, and speak indistinctly of
propositions and projectors.

The premises behind the BKS theorem can be formulated as follows:
\begin{enumerate}
\item[(a)]  {\it In an individual system each proposition $P_{{\bf u}_i}$
has a unique answer, }$0${\it \ or }$1$,{\it \ which is independent of which
other observables are being considered jointly (non-contextuality).}
\item[(b)]  {\it For each set of one dimensional projectors whose sum is the
unit matrix in the n-dimensional Hilbert space of the states of the system,
the answer to one and only one of the projectors is }$1${\it , and the
answers to the other }$n-1${\it \ are }$0${\it .}
\end{enumerate}
Following these rules, if the answers to the projector over a given vector
is 1, the answers to the projectors over all orthogonal vectors must be zero.

Let us consider the answers to the projectors over the following 9 sets of
orthogonal four-dimensional vectors: 
\begin{equation}
\label{e1}v\left( 0,0,0,1\right) +v\left( 0,0,1,0\right) +v\left(
1,1,0,0\right) +v\left( 1,-1,0,0\right) =1, 
\end{equation}
\begin{equation}
\label{e2}v\left( 0,0,0,1\right) +v\left( 0,1,0,0\right) +v\left(
1,0,1,0\right) +v\left( 1,0,-1,0\right) =1, 
\end{equation}
\begin{equation}
\label{e3}v\left( 1,-1,1,-1\right) +v\left( 1,-1,-1,1\right) +v\left(
1,1,0,0\right) +v\left( 0,0,1,1\right) =1, 
\end{equation}
\begin{equation}
\label{e4}v\left( 1,-1,1,-1\right) +v\left( 1,1,1,1\right) +v\left(
1,0,-1,0\right) +v\left( 0,1,0,-1\right) =1, 
\end{equation}
\begin{equation}
\label{e5}v\left( 0,0,1,0\right) +v\left( 0,1,0,0\right) +v\left(
1,0,0,1\right) +v\left( 1,0,0,-1\right) =1, 
\end{equation}
\begin{equation}
\label{e6}v\left( 1,-1,-1,1\right) +v\left( 1,1,1,1\right) +v\left(
1,0,0,-1\right) +v\left( 0,1,-1,0\right) =1, 
\end{equation}
\begin{equation}
\label{e7}v\left( 1,1,-1,1\right) +v\left( 1,1,1,-1\right) +v\left(
1,-1,0,0\right) +v\left( 0,0,1,1\right) =1, 
\end{equation}
\begin{equation}
\label{e8}v\left( 1,1,-1,1\right) +v\left( -1,1,1,1\right) +v\left(
1,0,1,0\right) +v\left( 0,1,0,-1\right) =1, 
\end{equation}
\begin{equation}
\label{e9}v\left( 1,1,1,-1\right) +v\left( -1,1,1,1\right) +v\left(
1,0,0,1\right) +v\left( 0,1,-1,0\right) =1. 
\end{equation}
There are 18 different vectors in (\ref{e1}--\ref{e9}). $S$ will denote this
set of vectors, and $P$ the set of the corresponding propositions. Our
state-independent version of the BKS theorem can be enunciated as follows:
\begin{quote}
{\it There is no set of answers satisfying }(a){\it \ and }(b){\it \ to the
set of propositions }$P$.
\end{quote}
The proof is straightforward: the sum of the right-hand sides of (\ref{e1}--%
\ref{e9}) is {\em odd}, whereas the sum of the left-hand sides is
necessarily {\em even}, because each answer appears twice. The previous
record \cite{5} involved 11 equations with 20 vectors, appearing either
twice or four times each.

The vectors in $S$ can be interpreted as pure spin states of a system of two
spin-$\frac{1}{2}$ particles. The 12 vectors in (\ref{e1}--\ref{e4}) are factorizable
(i.e., of the form $\left( a,b\right) ^{\left( 1\right) }\otimes \left(
c,d\right) ^{\left( 2\right) }$), and the corresponding projectors are
products of local observables. For instance, in the usual Pauli
representation, the (unnormalized) vector $\left( 1,-1,0,0\right) $
represents the state $\left| \sigma _z=+1\right\rangle ^{\left( 1\right)
}\otimes \left| \sigma _x=-1\right\rangle ^{\left( 2\right) }$, and its
corresponding projector represents the proposition: does the observable $%
\widehat{\sigma }_z^{\left( 1\right) }$ have a well-defined (hidden) value $%
+1$ {\em and} the observable $\widehat{\sigma }_x^{\left( 2\right) }$ a
well-defined (hidden) value $-1$?, with $v\left( 1,-1,0,0\right) =1$ if the
answer is ``yes'', and $v\left( 1,-1,0,0\right) =0$ otherwise.

The remaining six vectors in (\ref{e5}--\ref{e9}) are entangled, and the
corresponding propositions cannot be factorized in terms of local
observables. Each one can be expressed in terms of a pair of the observables 
$\widehat{\sigma }_z^{\left( 1\right) }\otimes \widehat{\sigma }_z^{\left(
2\right) }$, $\widehat{\sigma }_z^{\left( 1\right) }\otimes \widehat{\sigma }%
_x^{\left( 2\right) }$, $\widehat{\sigma }_x^{\left( 1\right) }\otimes 
\widehat{\sigma }_z^{\left( 2\right) }$ and $\widehat{\sigma }_x^{\left(
1\right) }\otimes \widehat{\sigma }_x^{\left( 2\right) }$ \cite{7,8}. For
instance, $\left( 1,-1,1,1\right) $ is an eigenvector of $\widehat{\sigma }%
_z^{\left( 1\right) }\otimes \widehat{\sigma }_x^{\left( 2\right) }$ and $%
\widehat{\sigma }_x^{\left( 1\right) }\otimes \widehat{\sigma }_z^{\left(
2\right) }$ with eigenvalues $-1$ and $+1$, respectively, and therefore can
be associated with the proposition: do the observables $\widehat{\sigma }%
_z^{\left( 1\right) }\otimes \widehat{\sigma }_x^{\left( 2\right) }$ {\em and%
} $\widehat{\sigma }_x^{\left( 1\right) }\otimes \widehat{\sigma }_z^{\left(
2\right) }$ have well-defined (hidden) values $-1$ {\it and} $+1$,
respectively?. Each of the equations (\ref{e5}--\ref{e8}) involves a pair of
these entangled vectors, whereas (\ref{e9}) involves four.

A state-independent BKS\ proof is said to be ``critical'' \cite{9} if it is
based on a set of propositions not having any subset also making possible a
state-independent proof. Peres' set of 24 vectors\footnote{%
The vectors in Peres' set \cite{7,8} can be geometrically interpreted as
vectors along the 24 directions that join the center of a four-dimensional
hypercube (tesseract) with the (pairwise opposite) centers of its 8
three-dimensional faces (cubes), the centers of the 24 two-dimensional
intersections of them (squares), and the 16 vertices. The sets of vectors in
several other BKS\ ``state-independent'' proofs have been nicknamed
according to their aspect (Kochen-Specker's 117--vector set \cite{2} is also
known as the ``cat's cradle'' \cite{10}, Peres' 33-vector set \cite{7,8} as
the ``quantum polyhedron'' \cite{11}, and Penrose's 40-vector set\ \cite
{9,12} as the ``magic dodecahedron'' \cite{13}); therefore we suggest naming
Peres' 24-vector set the ``quantum tesseract''.} is not critical; it
contains Kernaghan's 20-vector critical set \cite{5} and 95 other critical
sets of 20, plus our previous set $S$ and 15 other critical sets of 18. From
the definition of critical set there follows that none of these 18-vector
sets are contained in any of the 96 {\em critical} 20-vector sets, which
probably explains why they were not obtained previously. Peres' set does not
contain any subset with fewer than 18 vectors allowing of a
state-independent BKS\ proof. The assertions in this paragraph can be
checked by means of a computer program generalizing to dimension 4 the one
in ref. \cite{8}.

In \cite{4} we proved how, by increasing the number of vectors, we can go
from probabilistic demonstrations to state-specific and then to
state-independent ones. Here we will illustrate the reverse procedure,
showing how our 18-vector set $S$ contains subsets allowing of
state-specific and probabilistic BKS proofs.

Each vector in our set $S$ is orthogonal to seven other vectors in the set;
therefore, we can prepare the system in a state that assigns the answer $1$
to the projector over one of these vectors and the answer $0$ to the other
seven projectors over orthogonal vectors. For instance, if we prepare the
system in the singlet state, 
\begin{equation}
\label{e10}\left| \psi \right\rangle =\left( \left| +-\right\rangle -\left|
-+\right\rangle \right) /\sqrt{2}, 
\end{equation}
then, by definition 
\begin{equation}
\label{e11}v\left( 0,1,-1,0\right) =1, 
\end{equation}
and we can discard from eqs.~(\ref{e1}--\ref{e9}) the vector $\left(
0,1,-1,0\right) $ and those orthogonal to it, whose associated values are
zero, 
\begin{equation}
\label{e12}
\begin{array}{l}
v\left( 0,0,0,1\right) =v\left( 1,-1,-1,1\right) =v\left( 1,1,1,1\right)
=v\left( 1,0,0,-1\right) = \\ 
v\left( 1,1,1,-1\right) =v\left( -1,1,1,1\right) =v\left( 1,0,0,1\right) =0. 
\end{array}
\end{equation}
Therefore, only seven equations with 10 different vectors remain: 
\begin{equation}
\label{e13}v\left( 0,0,1,0\right) +v\left( 1,1,0,0\right) +v\left(
1,-1,0,0\right) =1, 
\end{equation}
\begin{equation}
\label{e14}v\left( 0,1,0,0\right) +v\left( 1,0,1,0\right) +v\left(
1,0,-1,0\right) =1, 
\end{equation}
\begin{equation}
\label{e15}v\left( 1,-1,1,-1\right) +v\left( 1,1,0,0\right) +v\left(
0,0,1,1\right) =1, 
\end{equation}
\begin{equation}
\label{e16}v\left( 1,-1,1,-1\right) +v\left( 1,0,-1,0\right) +v\left(
0,1,0,-1\right) =1, 
\end{equation}
\begin{equation}
\label{e17}v\left( 0,0,1,0\right) +v\left( 0,1,0,0\right) =1, 
\end{equation}
\begin{equation}
\label{e18}v\left( 1,1,-1,1\right) +v\left( 1,-1,0,0\right) +v\left(
0,0,1,1\right) =1, 
\end{equation}
\begin{equation}
\label{e19}v\left( 1,1,-1,1\right) +v\left( 1,0,1,0\right) +v\left(
0,1,0,-1\right) =1. 
\end{equation}
There is no way of assigning definite answers to the 10 propositions
appearing in these equations. The proof is the same as before: the sum of
the right-hand sides of (\ref{e13}--\ref{e19}) is {\em odd}, whereas the sum
of the left-hand sides is necessarily {\em even}, because each answer
appears twice.

Apparently this conclusion rests on the impossibility of unique answers to $%
10+8$ propositions: the 10 different ones in (\ref{e13}--\ref{e19}), plus
the one for the initial state (\ref{e11}) and the seven for orthogonal
vectors (\ref{e12}). But in fact we can justify eqs.~(\ref{e13}--\ref{e19})
without the assistance of (\ref{e12}), using the following argument \cite{3}%
: each subset of two or three vectors in the left-hand sides of (\ref{e13}--%
\ref{e19}) spans a subspace that contains the vector $\left( 0,1,-1,0\right) 
$ (we can check that this vector can be expressed as a linear combination of
the ones in each subset); therefore, even if the sums of the corresponding
projectors are not the $4\times 4$ unit matrix, the system is in an
eigenstate, with eigenvalue $1$, of each sum of projectors, and the sums of
the corresponding answers must be 1. In consequence, our state-specific
proof uses only 10 vectors (or $10\left( +1\right) $, if we also count the
initial state). The previous record \cite{3} involved seven equations with
13 (or $13\left( +1\right) $, if we include the initial state) different
eight-dimensional vectors, appearing either twice or four times each. Note
nevertheless that the state-specific proof in \cite{3} has the desirable
property of using only factorizable vectors (i.e., of the form $\left(
a,b\right) ^{\left( 1\right) }\otimes \left( c,d\right) ^{\left( 2\right)
}\otimes \left( e,f\right) ^{\left( 3\right) }$), as opposed to our
state-specific and state-independent proofs, or the state-independent one in 
\cite{5}.

Our state-specific BKS theorem cannot be interpreted as a contradiction
between quantum mechanics (QM in the following) and NCHV in terms of {\em %
local} measurements, because, although it is possible to prepare the system
in an entangled state (the singlet, in our previous choice), we cannot
eliminate the remaining five non-factorizable propositions, and still reach
a contradiction. For instance, our previous choice (\ref{e13}--\ref{e19})
contains the entangled state $\left( 1,1,-1,1\right) $, and the answer to
the corresponding propositions cannot be determined by means of a local
measurement on particle 1 and a local measurement on particle 2.

We will now obtain a probabilistic version of the BKS theorem using only
factorizable projectors, interpretable in terms of local measurements, and
showing the incompatibility between QM and local realistic theories. Other
correspondences between BKS' and Bell's theorems have been discussed in
literature \cite{14}.

Suppose we prepare (``preselect'') two spin-$\frac{1}{2}$ particles in the entangled
(but no ``maximally entangled'') Hardy state \cite{6}, 
\begin{equation}
\label{e20}\left| \eta \right\rangle =\left( \left| ++\right\rangle -\left|
+-\right\rangle -\left| -+\right\rangle \right) /\sqrt{3}. 
\end{equation}
Then, by definition 
\begin{equation}
\label{e21}v\left( 1,-1,-1,0\right) =1. 
\end{equation}
The answers to the projectors over any vector orthogonal to $\left(
1,-1,-1,0\right) $\footnote{%
This vector does not belong to $S$ nor to Peres' 24-vector set. It can be
geometrically interpreted as the direction that joins the centers of a pair
of opposite edges of a tesseract (see previous footnote); the other 15
directions joining the centers of the remaining opposite edges also
represent Hardy states \cite{6}.}, must be zero; in particular, 
\begin{equation}
\label{e22}v\left( 0,0,0,1\right) =v\left( 1,1,0,0\right) =v\left(
1,0,1,0\right) =0. 
\end{equation}
Let us assume that a subsequent measurement (``postselection'') finds the
system in the state 
\begin{equation}
\label{e23}\left| \varphi \right\rangle =\left| \sigma _x=+1\right\rangle
^{\left( 1\right) }\otimes \left| \sigma _x=+1\right\rangle ^{\left(
2\right) } 
\end{equation}
(this is possible because $\left\langle \varphi \right. \left| \eta
\right\rangle \neq 0$); then, 
\begin{equation}
\label{e24}v\left( 1,1,1,1\right) =1. 
\end{equation}
In the individual systems postselected in state (\ref{e23}), the answer to
all propositions over vectors orthogonal to $\left( 1,1,1,1\right) $ is $0$;
in particular 
\begin{equation}
\label{e25}v\left( 1,-1,0,0\right) =v\left( 1,0,-1,0\right) =0. 
\end{equation}
Replacing (\ref{e22}) and (\ref{e25}) in (\ref{e1}) and (\ref{e2}) leads to 
\begin{equation}
\label{e26}v\left( 0,1,0,0\right) =v\left( 0,0,1,0\right) =1. 
\end{equation}
But $\left( 0,1,0,0\right) $ and $\left( 0,0,1,0\right) $ are orthogonal,
and therefore the answers to the corresponding propositions cannot both be
1: we have reached a contradiction.

This probabilistic demonstration of the BKS theorem uses $7\,\left(
+2\right) $ vectors (seven in (\ref{e22},\ref{e25},\ref{e26}) plus the
states $\eta $ and $\varphi $). The term ``probabilistic'' follows from the
fact that preparing the system in the initial state $\eta $ (i.e.,
preselecting $\eta $) gives only a non-zero probability of finding the
system in the final state $\varphi $ (i.e., of postselecting $\varphi $),
not a certainty.

Now we are going to show how this result relates to Hardy's nonlocality
theorem \cite{6} . Note that all the vectors involved in the previous proof
are factorizable, with the exception of the initial state $\eta $. The
answer to a factorizable proposition can be expressed in terms of the
answers to the corresponding factors: 
\begin{equation}
\label{e27}v\left[ \left( a,b\right) ^{\left( 1\right) }\otimes \left(
c,d\right) ^{\left( 2\right) }\right] =1\Leftrightarrow v\left( a,b\right)
^{\left( 1\right) }=v\left( c,d\right) ^{\left( 2\right) }=1, 
\end{equation}
\begin{equation}
\label{e28}v\left[ \left( a,b\right) ^{\left( 1\right) }\otimes \left(
c,d\right) ^{\left( 2\right) }\right] =0\Leftrightarrow v\left( a,b\right)
^{\left( 1\right) }\times v\left( c,d\right) ^{\left( 2\right) }=0. 
\end{equation}
In particular, if we preselect the state $\eta $, 
\begin{equation}
\label{e29}v\left( 1,-1,-1,0\right) =1\Rightarrow v\left( 1,1,0,0\right)
=0\Leftrightarrow v\left( 1,0\right) ^{\left( 1\right) }\times v\left(
1,1\right) ^{\left( 2\right) }=0. 
\end{equation}
Similarly, postselecting $\varphi $ (i.e., $v[\left( 1,1\right) ^{\left(
1\right) }\otimes \left( 1,1\right) ^{\left( 2\right) }]=1$) implies, using (%
\ref{e27}), 
\begin{equation}
\label{e30}v\left( 1,1\right) ^{\left( 1\right) }=1, 
\end{equation}
\begin{equation}
\label{e31}v\left( 1,1\right) ^{\left( 2\right) }=1. 
\end{equation}
Then, (\ref{e29},\ref{e31}) imply 
\begin{equation}
\label{e32}v\left( 1,0\right) ^{\left( 1\right) }=0. 
\end{equation}
If we use premises (a) and (b) in the two-dimensional spin space of the
first particle, from (\ref{e32}) we conclude that 
\begin{equation}
\label{e33}v\left( 0,1\right) ^{\left( 1\right) }=1. 
\end{equation}
The answers (\ref{e21},\ref{e31},\ref{e33}) correspond in QM terms to the
following value for the conditional probability of finding $\sigma
_z^{\left( 1\right) }=-1$ in a system prepared in the state $\eta $, {\it if}
$\sigma _x^{\left( 2\right) }=+1$; 
\begin{equation}
\label{e34}P_\eta \left( \sigma _z^{\left( 1\right) }=-1\left| \sigma
_x^{\left( 2\right) }=+1\right. \right) =1. 
\end{equation}
If we interchange the roles of particles 1 and 2, a similar reasoning leads
us to 
\begin{equation}
\label{e35}P_\eta \left( \sigma _z^{\left( 2\right) }=-1\left| \sigma
_x^{\left( 1\right) }=+1\right. \right) =1. 
\end{equation}
Eq.~(\ref{e21}) and the first part of (\ref{e22}) ($v\left( 0,0,0,1\right) =0
$) translate into 
\begin{equation}
\label{e36}P_\eta \left( \sigma _z^{\left( 1\right) }=-1,\sigma _z^{\left(
2\right) }=-1\right) =0. 
\end{equation}
Finally, the fact that the system can be postselected in the state $\varphi $%
, used to obtain (\ref{e24}), means that 
\begin{equation}
\label{e37}P_\eta \left( \sigma _x^{\left( 1\right) }=+1,\sigma _x^{\left(
2\right) }=+1\right) >0. 
\end{equation}
Eqs.~(\ref{e34}--\ref{e37}) translate into QM terms the set of answers used
in our probabilistic BKS theorem. If we assume that particles 1 and 2 are
localized in two spacelike separated regions (this assumption was not
necessary for the BKS theorem), eqs.~(\ref{e34}--\ref{e37}) are just those
in Hardy's nonlocality theorem, which we can summarize as follows:
\begin{quote}
Let us consider a system of two spacelike separated particles prepared in
the spin state $\eta $, and suppose that we accept EPR's sufficient
condition for existence of elements of reality \cite{15}. In those
individual systems in which $\sigma _x^{\left( 2\right) }=+1$, and $\sigma
_x^{\left( 1\right) }=+1 $ (condition that can be fulfilled because of (\ref
{e37})), eqs.~(\ref{e34},\ref{e35}) imply that we can jointly infer two
elements of reality, $\sigma _z^{\left( 1\right) }=-1$, $\sigma _z^{\left(
2\right) }=-1$. But these results can be never be obtained in a joint
measurement in the state $\eta $, because of (\ref{e36}): QM and elements of
reality are not compatible, {\it q.e.d}.
\end{quote}
We could also establish the inverse correspondence: eqs.~(\ref{e34}--\ref
{e37}) in Hardy's theorem can be translated into a set of definite answers
to propositions that proves our probabilistic BKS theorem; we omit the
details for brevity.

In summary: we have found sets of four-dimensional real vectors that allow
us to develop state-independent, state-specific and probabilistic BKS
proofs, illustrating the relations between these three versions of the
theorem. In the first two cases, our sets are the most economical yet, in
terms of vectors used, in any dimension. On the other hand, the
probabilistic proof shows the same kind of contradiction as Hardy's theorem,
and suggest an algebraic reading of it.\\

We would like to thank Gabriel \'Alvarez and Jos\'e Luis Cereceda for
reading this paper and making valuable comments.

\end{document}